\documentclass[12pt]{article}
\usepackage[cp1251]{inputenc}

\usepackage{amssymb}
\usepackage{amsmath}
\usepackage{latexsym}
\usepackage{yfonts}

\catcode`@=11
\renewcommand{\section}{\@startsection{section}{1}{0pt}{\medskipamount}
{\medskipamount}{\large\bf}}
\numberwithin{equation}{section}
\catcode`@=12
\def\beq{\begin{eqnarray}}    %%%  begequation/eqnarray
\def\eeq{\end{eqnarray}}      %%%  endequation/eqnarray

\def\sdet{\,\mbox{sdet}\,}              %%% superdeterminant

\def\={\ =\ }

\begin{document}

\begin{center}

{\Large\bf Physical quantities  and arbitrariness
in resolving quantum master equation
}

\vspace{18mm}

{\large Igor A. Batalin$^{(a,b)}\footnote{E-mail:
batalin@lpi.ru}$\;, Peter M. Lavrov$^{(b, c)}\footnote{E-mail:
lavrov@tspu.edu.ru}$\; }

\vspace{8mm}

\noindent ${{}^{(a)}}$
{\em P.N. Lebedev Physics Institute,\\
Leninsky Prospect \ 53, 119991 Moscow, Russia}

\vspace{3mm}

\noindent  ${{}^{(b)}} ${\em
Tomsk State Pedagogical University,\\
Kievskaya St.\ 60, 634061 Tomsk, Russia}

\vspace{3mm}

\noindent  ${{}^{(c)}} ${\em
National Research Tomsk State  University,\\
Lenin Av.\ 36, 634050 Tomsk, Russia}

\vspace{20mm}

\begin{abstract}
\noindent
By proceeding with the idea that the presence of physical (BRST invariant) extra factors
in the path integral is equivalent to taking into account explicitly the arbitrariness in
resolving the quantum master equation, we consider the field-antifield quantization
procedure both with the Abelian and the non-Abelian gauge fixing.
\end{abstract}

\end{center}

\vfill

\\

\newpage

\section{Introduction}
\vspace{3mm}

Let $W$ be a quantum master action, and let $\mathcal{ O }$ be a physical
(BRST invariant)
quantity. Then, the product, $\exp\{ \frac{ i }{ \hbar }  W \} \mathcal{ O }$,
resolves the quantum
master equation, as well. Thus, the presence of physical extra factors
in the path integral
is equivalent to taking into account explicitly the arbitrariness
in resolving the quantum
master equation.

     In the present paper, by proceeding with the above simple idea,
we consider the field-antifield  quantization procedure
\cite{BV,BV1,VLT,LT,BV2,BB1,BLT-EPJC,BL1},
both with the Abelian and
the non-Abelian gauge fixing.

     In the case of a non-Abelian gauge fixing, a rather non-trivial measure
in the path integral
is given by the so-called "square root formula" \cite{BT,BBD-2006,Kh}.
There are two aspects
in dependence
of the measure on the elements of gauge arbitrariness. Formally, the measure does depend
both on the gauge fixing functions themselves, and on the functions that complement the
latter gauge fixing functions to constitute together an invertible reparametrization
as for
the total set of the field-antifield variables.

     In \cite{BBD-2006}, it has been shown that the path integrand is actually independent
of the complement
functions. Furthermore, it has then been shown in \cite{BBD-2006}
that the whole path integral
is actually
independent on-shell of the non-Abelian gauge fixing functions as well.
In the present paper,
we use essentially these results.
\\

\section{Abelian gauge-fixing}
\vspace{3mm}

Let us consider the equation for a physical quantity
$\exp\big\{ \frac{ i }{ \hbar } X \big\}$,
\beq
\label{a1}
\sigma  \exp\Big\{ \frac{ i }{ \hbar } X \Big\}  =  0,  %   (1)
\eeq
where
\beq
\label{a2}
\sigma  =:  \frac{ \hbar }{ i }  \exp\Big\{ - \frac{ i }{ \hbar } W \Big\}
\overrightarrow{ \Delta }  \exp\Big\{ \frac{ i }{ \hbar } W \Big\}  =
\frac{ \hbar }{ i }  \Delta  +  {\rm ad}( W ),   %   (2)
\eeq
is the total BRST-operator,  and   $W$ is a solution
to the quantum master equation,
\beq
\label{a3}
\Delta  \exp\Big\{ \frac{ i }{ \hbar } W \Big\}  =  0,   %  (3)
\eeq
with the $\Delta$ being the odd Laplacian,
\beq
\label{a4}
\Delta  =:  (-1)^{ \varepsilon_{A} }
\frac{ \partial }{ \partial \Phi^{A} }  \frac{ \partial }{ \partial \Phi^*_{A} },   %        (0.4)
\eeq
\beq
\label{a5}
\varepsilon( \Phi^{A} )  =:  \varepsilon_{A}  =:  \varepsilon( \Phi^*_{A} )  +  1. %(0.5)
\eeq
 It follows from  (\ref{a1}), (\ref{a2}) that the sum  $( W + X )$  does also satisfy
 the equation  (\ref{a3}).
Therefore, it there holds
\beq
\label{a6}
\exp\Big\{ \frac{ i }{ \hbar } ( W + X ) \Big\}  =  \exp\{ [ \Delta, F ] \}
\exp\Big\{\frac{ i }{ \hbar } W \Big\},    %  (4)
\eeq
as the operator  $\exp\{ [ \Delta, F] \}$, with an  arbitrary Fermion operator
 $F$, does act transitively on the set of solutions to the equation
(\ref{a3}) under the suitable boundary conditions.
In turn,  the solution  (\ref{a6}) rewrites in the form
\beq
\label{a7}
\exp\Big\{ \frac{ i }{ \hbar } X \Big\}  =  \exp\Big\{ - \frac{ i }{ \hbar } W \Big\}
\exp\{ [ \Delta, F ] \}  \exp\Big\{ \frac{ i }{ \hbar } W \Big\}  =
\exp\Big\{ \frac{ i }{ \hbar} [ \sigma, F' ] \Big\} \cdot  1 ,  %    (5)
\eeq
where  $F'$ is the transformed Fermion operator $F$,
to be considered as an arbitrary one, as well,
\beq
\label{a8}
F'  =:  \exp\Big\{ - \frac{ i }{ \hbar } W \Big\}  F  \exp\Big\{ \frac{ i }{ \hbar }
W \Big\}.    %   (6)
\eeq
 Being the  $F$ a function, then $F' = F$ is a function as well.
It is just the operator
(\ref{a8}) that might depend on  the so-called "composite operators", as there is
no  other arbitrariness in the solution
(\ref{a7}).
Thereby, the expression  for the generating functional
has the form
\beq
\nonumber
\mathcal{ Z }[ J ]  &=:& \! \int\! [ D \Phi ] [ D \Phi^* ] [D  \Lambda ]
\exp\Big\{  \frac{ i }{ \hbar }  \Big[  W  +  X  +  J_{A}  \Phi^{A}  +\\
\label{a9}
&&\qquad\qquad\qquad\qquad+\Big(  \Phi^*_{A}  -  \Psi( \Phi)
\frac{ \overleftarrow{
\partial }}{ \partial \Phi^{A} } \Big)\Lambda^A \Big]  \Big\}, \!\!  %   (7)
\eeq
with  $\Psi( \Phi)$ being the gauge fixing Fermion, and
$\varepsilon( \Lambda^{A} )  =:  \varepsilon_{A}  +  1$. Obviously, the
effective action corresponding to the generating functional (\ref{a9})
is gauge independent
on-shell $J = 0$.
Indeed, by making in (\ref{a9}) the BRST transformation,
\beq
\label{a10}
\delta \Phi^{A}  &=:&  \Lambda^{A}  \mu,   \\   %   (0.9)
\label{a11}
\delta \Phi^*_{A}  &=:&
\mu
\Big(( W + X )\frac{ \overleftarrow{\partial }}{\partial \Phi^{A} }\Big),  \\ %   (0.10)
\label{a12}
\delta \Lambda^{A}  &=:&  0,  \\   %   (0.11)
\label{a13}
\varepsilon( \mu )  &=:&  1,      %     (0.12)
\eeq
with $\mu   =  {\rm const}$, we get the Ward identity
\beq
\label{a14}
J_{A}  \big\langle(  \Phi^{A},  ( W  +  X )  )\big\rangle  =  0,    %  (0.13)
\eeq
where the $\langle \cdots  \rangle$ means the functional average value with the
weight functional in (\ref{a9}), and the $( \cdot, \cdot )$ stands for the antibracket,
\beq
\nonumber
(A,B)&=:&(-1)^{\varepsilon(A)}[[\Delta,A],B]\cdot 1=\\
\label{a15}
&=& A\frac{ \overleftarrow{\partial }}{\partial \Phi^{C} }
\frac{ \overrightarrow{\partial }}{\partial \Phi^*_{C} }B -
(A \leftrightarrow B)(-1)^{(\varepsilon(A)+1)(\varepsilon(B)+1)},
\eeq
as for two arbitrary functions $A,B$. The latter antibracket satisfies
 the differentiation property,
\beq
\label{a16}
\Delta (A,B)=(\Delta A,B)-(A,\Delta B)(-1)^{\varepsilon(A)} ,
\eeq
and  the non-polarized Jacobi identity,
\beq
\label{a17}
((B,B),B)=0,\quad\varepsilon(B)=0.
\eeq
In terms of the antibracket (\ref{a15}) the quantum master equation (\ref{a3})
rewrites naturally in its quadratic form
\beq
\label{a18}
\frac{1}{2}(W,W)+\frac{\hbar}{i}\Delta W=0,
\eeq
so that we have for $X$ the respective quadratic equation of the
form,
\beq
\label{a18a}
\frac{1}{2}  ( X, X )  +  \sigma  X  =  0,   %  (2.19)
\eeq
with $\sigma$ given in (\ref{a2}).

On the other hand, by choosing in (\ref{a10})-(\ref{a13}),
\beq
\label{a19}
\mu  =:  \frac{ i }{ \hbar }  \delta \Psi( \Phi),      %   (0.14)
\eeq
we arrive at  the gauge independence on-shell,
\beq
\label{a20}
\delta_{ \Psi }  \mathcal{ Z }( J  =  0 )  =  0.   %      (0.15)
\eeq

Finally, we give a more explicit form of the solution
(\ref{a7}) in the case when  $F' = F$ is a function,
\beq
\nonumber
\exp\Big\{ \frac{ i }{ \hbar } X \Big\}  &=&
\exp\{ ( E( - {\rm ad}( F ) ) \Delta F ) \}\times\\
\label{a21}
&&\times
\exp\Big\{ - \frac{ i }{ \hbar } W \Big\} \exp\{ - {\rm ad}( F ) \}
\exp\Big\{ \frac{ i }{ \hbar } W \Big\},    %   (8)
\eeq
where the notations
\beq
\label{a22}
E( Z )  =:  \frac{  \exp\{ Z \}  -  1  }{ Z }    %  (9)
\eeq
and
\beq
\label{a23}
{\rm ad}(A)B=:(A,B)
\eeq
are used.
\\

\section{Non-Abelian gauge-fixing}
\vspace{3mm}

Let us generalize the generating functional (\ref{a9}) to cover the case of
non-Abelian gauge fixing functions $G_{A}$ that satisfy the antibracket
involution relations,
\beq
\label{b1}
( G_{A}, G_{B} )  =  \mathcal{ U }_{AB}^{C}  G_{C},  %    (1.1)
\eeq
together with the conditions,
\beq
\label{b2}
\varepsilon( G_{A} )  =  \varepsilon_{A}  +  1,   %     (1.2)
\eeq
the even matrix,
\beq
\label{b3}
\left \| \frac{ \partial G_{A} }{ \partial \Phi^*_{B} } \right \|,  %     (1.3)
\eeq
is invertible.
  Let the functions $F^{A}$ that satisfy
\beq
\label{b4}
\varepsilon( F^{A} )  =  \varepsilon_{A},     %      (1.4)
\eeq
do complement the gauge fixing functions $G_{A}$ in the sense that the
reparametrization,
\beq
\label{b5}
\Phi^{A}, \Phi^*_{A}  \;  \Rightarrow   \; F^{A}, G_{A},     %   (1.5)
\eeq
is invertible, and let $\mathcal{ J }$ be the super-Jacobian of the latter
reparametrzation (\ref{b5}).
   Let us consider the generating functional given by the "square root
formula" \cite{BBD-2006},
\beq
\nonumber
\mathcal{ Z }[J]  &=:&  \int [ D \Phi ] [ D \Phi^* ] [ D \Lambda ]
\sqrt{ \mathcal{ J }  \sdet ( ( F^{A}, G_{B} ) )  }\times\\
\label{b6}
&&\times\exp\Big\{ \frac{ i }{ \hbar } [ W  +  X  +  J_{A} \Phi^{A}  +
G_{A} \Lambda^A] \Big\}.    %    (1.6)
\eeq
It has been shown in \cite{BBD-2006}  that the integrand in (\ref{b6})
is independent of $F^{A}$.
Furthermore, it has then been shown in \cite{BBD-2006} that the path integral (\ref{b6})
is independent of $G_{A}$ as well on-shell $J = 0$.

The factors inside the square root in the integrand in (\ref{b6}) can be presented
in their integral form
\beq
\label{b7}
\mathcal{ J }  =:  \int [ D \bar{ C } ] [ D C ]
\exp\Big\{  \frac{ i }{ \hbar } (  \bar{ C }_{A}  F^{A}  +  \bar{C}^{A}  G_{A}  )
\Big(  \frac{ \overleftarrow{ \partial }}{ \partial \Phi^{B} }  C^{B}  +
\frac{ \overleftarrow{ \partial }}{ \partial \Phi^*_{B} }  C_{B}  \Big)  \Big\},    %     (1.7)
\eeq
\beq
\label{b8}
\sdet( ( F^{A}, G_{B} ) )  =:  \int [ D \bar{ B } ] [ D B ]
\exp\Big\{  \frac{ i }{ \hbar }  \bar{ B }_{A}  ( F^{A}, G_{C} )  B^{ C}  \Big\},   %      (1.8)
\eeq
where the respective Grassmann parities are
\beq
\label{b9}
\varepsilon( \bar{ C }_{A} )  &=:&  \varepsilon_{A}  +  1,  \\%     (1.9)
\label{b10}
\varepsilon( \bar{ C }^{A} )  &=:&  \varepsilon_{A},  \\%     (1.10)
\label{b11}
\varepsilon( C^{A} )  &=:&  \varepsilon_{A}  +  1,  \\%      (1.11)
\label{b12}
\varepsilon( C_{A} )  &=:&    \varepsilon _{A},   \\%     (1.12)
\label{b13}
\varepsilon( \bar{ B}_{A} )  &=:&  \varepsilon_{A} ,  \\%    (1.13)
\label{b14}
\varepsilon( B^{A} )  &=:&  \varepsilon_{A} .     %   (1.14)
\eeq

Now, by multiplying the integral (\ref{b7}) by the one (\ref{b8}),
consider, in their product,
the "BRST" transformations,
\beq
\label{b15}
\delta C^{A}  &=:&  ( \Phi^{A}, G_{D} )  B^{D}  \mu,   \\%   (3.9)
\label{b16}
\delta C_{A}  &=:&  ( \Phi^*_{A}, G_{D} )  B^{D} \mu,  \\% (3.10)
\label{b17}
\delta \bar{ B }_{A}  &=:&  \mu  \bar{ C }_{A},   \\%          (3.11)
\label{b18}
\varepsilon( \mu )  &=:&  1.   %        (3.12)
\eeq
It follows that the product of the integrals (\ref{b7}) and (\ref{b8})
 is invariant on-shell, $G_{A} = 0$,
under the transformations (\ref{b15})-(\ref{b18}),
as for the case of $\mu = {\rm const}$.

On the other hand, when choosing
\beq
\label{b19}
\mu  =: \frac{ i }{ \hbar } \bar{ B }_{A}  \delta F^{A}
\Big(  \frac{ \overleftarrow{ \partial } }{ \partial \Phi^{D} }  C^{D}  +
 \frac{ \overleftarrow{ \partial } }{ \partial \Phi^*_{D} }  C_{D}  \Big),    %     (3.13)
\eeq
in  (\ref{b15})-(\ref{b18}), one reproduces exactly the infinitesimal change
\beq
\label{b20}
F^{A}  \; \rightarrow \;  F^{A}  +  \delta F^{A},   %  (3.14)
\eeq
in the product of the integrals (\ref{b7}) and (\ref{b8}).

Thus, we have shown that the integrand in the path integral (\ref{b6})
is independent of $F^{A}$.
Now, let us choose $F^{A}  =  \Phi^{A}$. Then, it follows immediately
that the path integral (\ref{b6})
reduces to its form (\ref{a9}) as for the case of the Abelian gauge fixing.

To compare for \cite{BBD-2006}, here we have used the same "BRST" transformations
(\ref{b15})-(\ref{b18}),
although without explicit use of the intrinsic mini-version of
the master equation as for the
$BC$- ghost system.
\\

\section*{Acknowledgments}
\noindent  The work of I. A. Batalin is supported in part by the
RFBR grant 17-02-00317. The work of P. M. Lavrov is
supported by the Ministry of Education and Science of
Russian Federation, grant  3.1386.2017.
\\


\begin {thebibliography}{99}
\addtolength{\itemsep}{-8pt}

\bibitem{BV}
I. A. Batalin, G. A. Vilkovisky, {\it Gauge algebra and quantization},
Phys. Lett. B {\bf 102} (1981) 27 - 31.

\bibitem{BV1}
I. A. Batalin, G. A. Vilkovisky, {\it Quantization of gauge theories with linearly
dependent generators}, Phys. Rev. D {\bf 28} (1983) 2567 - 2582.

\bibitem{VLT}
B. L. Voronov, P. M. Lavrov and I. V. Tyutin,
{\it Canonical transformations and gauge dependence
in general gauge theories}, Sov. J. Nucl. Phys. {\bf 36} (1982) 292.

\bibitem{LT}
P. M. Lavrov, I. V. Tyutin,
{\it Gauge theories of general form},
Sov. Phys. J. {\bf 25} (1982) 639 - 641.

\bibitem{BV2}
I. A. Batalin and G. A. Vilkovisky,
{\it Closure of the gauge algebra,
generalized Lie algebra equations and Feynman rules},
Nucl. Phys. B {\bf 234} (1984) 106 - 124.

\bibitem{BB1}
I. A. Batalin, K. Bering, {\it Gauge independence
in a higher-order Lagrangian formalism via change of  variables in the path integral},
Phys. Lett. B {\bf 742} (2015) 23 - 28.

\bibitem{BLT-EPJC}
I. A. Batalin, P. M. Lavrov, I. V. Tyutin,
{\it Finite anticanonical transformations in field-antifield formalism},
Eur. Phys. J. C {\bf 75} (2015) 270-1-16.

\bibitem{BL1}
I. A. Batalin, P. M. Lavrov, {\it Closed description of arbitrariness
in resolving quantum master equation},
Phys. Lett. B {\bf 758} (2016) 54 - 58.

\bibitem{BT}
I.A. Batalin, I.V. Tyutin,
{\it On the multilevel field - antifield formalism with the most general
Lagrangian hypergauges},
 Mod. Phys. Lett. A {\bf 9} (1994) 1707-1716.

\bibitem{BBD-2006}
I. A. Batalin, K. Bering, P. H. Damgaard,
{\it On generalized gauge-fixing in the field-antifield formalism},
Nucl. Phys. B {\bf 739} (2006) 389 - 440.

\bibitem{Kh}
O. M. Khudaverdian, private communication.

\end{thebibliography}
\end{document}